
                           \hsize=5  in\vsize=7.5in
\parindent=0in\parskip=0in
\null
{\it to appear in: Proc. of the Oviedo Symposium 1993 on "Fundamental
     Problems in Quantum Physics"; eds.: M.Ferrero and A. van der Merwe,
     (Kluwer); also gr-qc/9410037}\hfill
\bigskip\bigskip
\bf
CONTINUOUSLY DIAGONALIZED DENSITY OPERATOR\hfill\break
OF OPEN SYSTEMS\hfill
\bigskip\bigskip
\line{\hskip .8truein Lajos Di\'osi\hfill}
\bigskip
\it
\line
{\hskip .8truein KFKI Research Institute for Particle and Nuclear Physics
\hfill}
\line
{\hskip .8truein H-1525 Budapest 114, POB 49, Hungary\hfill}
\bigskip\bigskip
\rm
We showed several years ago that the density operator of Markovian open
systems can be diagonalized continuously in time. The resulting pure
state jump processes correspond to quantum trajectories proposed
in recent quantum optics calculations or, at fundamental level,
to exact consistent histories.
\bigskip
Key words:
open systems, jump process, quantum trajectories, consistent histories.\hfill
\bigskip\bigskip
\parindent=.5truein

The quantum state of an open quantum system --- actually a subsystem of a
closed one --- cannot be described by deterministically evolving state
vectors $\psi$. A given pure initial state
$$
\rho=P~~~~\bigl(=\psi\psi^\dagger\bigr)
\eqno(1)$$
turns into {\it mixed} state immediately after the system has interacted with
its environment:
$$
\rho\rightarrow\rho^\prime\not=U\rho U^\dagger.
\eqno(2)$$
Consider the {\it diagonalization} of the mixed density operator
$\rho^\prime:$
$$
\rho^\prime=\sum_n p_n^\prime P^\prime_n.
\eqno(3)$$
It is well known that the eigenstates $P^\prime_n$ can be interpreted as
possible pure states of the system:
$$
\rho^\prime=P^\prime_n~,
\eqno(4)$$
with the corresponding probability $p^\prime_n$;\  $(n=1,2,\dots)$.
In such a way a pure state representation can be maintained even after
the interaction. If, as usual, interactions occur repeatedly then the
{\it stochastic jump} (4) must be introduced repeatedly
after each interaction.

We restrict ourself to  open systems of permanent
idealized interaction with the environment, resulting in
{\it Markovian} evolution equation for the density operator:
$$
d\rho/dt= {\cal L}\rho\equiv -i[H,\rho]+\dots~,
\eqno(5)$$
where $\dots$ stands for terms representing non-unitary evolution.
These terms make any given initial pure state
$$
\rho(t_0)=P(t_0)
\eqno(6)$$
mixed during an arbitrary (short) period $\epsilon$.
For time $t_1=t_0+\epsilon$,
the resulting mixed density operator
$$
\rho(t_1)=e^{\epsilon\cal L}P(t_0)
\eqno(7)$$
may be diagonalized:
$$\rho(t_1)=\sum_n p_n(t_1)P_n(t_1).
\eqno(8)$$
So, at $t=t_1+0$, we can restore the system's pure state as
$$
\rho(t_1+0)=P_{n_1}(t_1),
\eqno(9)$$
with probability $p_{n_1}(t_1);$\  $(n_1=1,2,3,\dots)$.
Again, at $t_2=t_1+\epsilon$, we get mixed density operator
$$
\rho(t_2)=e^{\epsilon\cal L}P_{n_1}(t_1),
\eqno(10)$$
and we diagonalize it:
$$
\rho(t_2)=\sum_n p_n(t_2)P_n(t_2).
\eqno(11)$$
At $t=t_2+0$, we restore pure states
$$
\rho(t_2+0)=P_{n_2}(t_2),
\eqno(12)$$
with probability $p_{n_2}(t_2);$\  $(n_2=1,2,3,\dots)$.
And so on, for
$t_3=t_0+3\epsilon,\dots,t_\nu=t_0+\nu\epsilon$.

Accordingly, one has constructed a stochastic process for the
{\it conditional} quantum state $\rho(t)$ of the open system.
The conditional $\rho(t)$ jumps into
a pure state at times
$t_0,t_1,\dots,t_\nu$, while it is getting slightly mixed between
the jumps.
In average, the process recovers the {\it unconditional} density operator
of the system, satisfying the ensemble evolution equation (5).

If we let $\epsilon$ go to zero (at $\nu\epsilon=const.$) then
the conditional state $\rho(t)$ will be pure all the time.
We proved the existence of this limit in 1985 [1].
The limiting process is a generalized
{\it Poisson} (jump) {\it process} for the pure state
$$
\rho(t)\equiv \psi(t)\psi^\dagger(t).
\eqno(13)$$
The analytic expressions of the pure state jump process contain two
nonlinear operators: the {\it frictional} Hamiltonian $H_\psi$ and the
{\it transition rate} operator $W_\psi$; see [2] and [3].
Then, the pure state satisfies the frictional Schr\"odinger equation
$$
{d\over dt}\psi(t)=-iH_{\psi(t)}\psi(t)
\eqno(14)$$
for most of the time, apart from the
discrete orthogonal jumps
$$
\psi(t+0)=\psi_n(t)
\eqno(15)$$
to the {\it n}th eigenstate of the current transition rate operator
$W_{\psi(t)}$. The transition rate of the jump is equal to the {\it n}th
eigenvalue $w_n(t)$ of $W_{\psi(t)}$.

The stochastic average of the pure state density operator (13) recovers
the unconditional density operator and satisfies the ensemble evolution
equation (5).

References 2 and 3 show jump process' equations for the general
evolution equation (5). Here we consider the simplest Lindblad [4]
structure:
$$
d\rho/dt=-i[H,\rho]+F\rho F^\dagger-{1\over2}\{F^\dagger F,\rho\}
\eqno(16)$$
where $F$ is the only Lindblad generator. Let us relate it to concrete
(open) physical systems and enlist typical cases:

---in spontaneous emission, $F=const.\times\vert0><1\vert,$

---in damped cavity oscillation, $F=const.\times a,$

---in pumped laser, $F=const.\times a^\dagger a,$

---in Brownian motion, $F=const.\times q+i~const.\times p,$

---in Stern-Gerlach apparatus, $F=const.\times\sigma_z.$\hfill\break
The coupling constants set the strenghts of the environmental interactions.

As can be be shown [2,3],
the frictional Schr\"odinger equation (14) takes the form
$$\eqalign{
{d\psi\over dt}=-iH\psi&+{1\over2}\left(<F^\dagger>F-H.C.\right)\psi\cr
        &-{1\over2}\left(F^\dagger-<F^\dagger>\right)\left(F-<F>\right)\psi
         +{1\over2}\Delta_F^2\psi\cr},
\eqno(17)$$
where $\Delta_F^2=<F^\dagger F>-<F^\dagger><F>$.
The above deterministic evolution of the
state vector happens to be interrupted by the orthogonal jumps (15)
which, as follows from [2] and [3], take the form
$$
\psi\rightarrow {1\over\sqrt{w}}\left(F-<F>\right)\psi.
\eqno(18)$$
The rate $w$ of the above transition, appearing in the normalization factor,
is just equal to the
Lindblad generator's quantum spread $\Delta_F$ in the current state $\psi$.
(If I had more than one Lindblad generator in the evolution equation (16)
then more
than one outcome would exist for the jump, each with its partial transition
rate $w_n$; see in [2,3].)

In 1992 Dalibard {\it et al.} [5] considered the simple quantum optical Bloch
equation which corresponds to our evolution equation (16) with the
special choice
$F=\sqrt{\Gamma}\vert0><1\vert$. The authors construct a pure state
jump process, similar to ours. Their frictional Schr\"odinger equation reads
$$
{d\psi\over dt}=-iH\psi-{1\over2}F^\dagger F\psi+{1\over2}<F^\dagger F>\psi
\eqno(19)$$
while their jump is
$$
\psi\rightarrow {1\over\sqrt{w}} F\psi.
\eqno(20)$$
The jump rate $w$ is equal to $<F^\dagger F>$.

No doubt, the jump process (19,20) of Dalibard {\it et al.} is
a bit simpler
to implement on a computer as compared to our jump process (17,18).
{}From theoretical
point of view, however, the orthogonal jump process  which we have
obtained by the continuous diagonalization of the density operator seems
to be more justified. First of all, as shown in [1], the orthogonal
jump process
is {\it observable}. Let us consider the Hermitian operator
$$
{\cal O}={1\over w}(F-<F>\psi \psi^\dagger(F^\dagger-<F^\dagger>)
\eqno(21)$$
which is actually $1/w$ times the transition rate operator itself. Its
eigenvalues are 0 or 1.
It is shown in [1] that a {\it continous observation} of ${\cal O}$ is
possible. No quantum Zeno paradox will enter because the Markovian
dynamics represents stonger effects than the continuous observation.
This latter leads just to the orthogonal jump process
(17,18). For most of time, the observed value of ${\cal O}$ (21) is $0$ while
the state vector obeys to the nonlinear Schr\"odinger equation (17).
For some random instants, however, the observed value may be $1$: then
the state vector has just performed the orthogonal jump (18).

That the above observability of the jump process is fundamental
has got new support recently. There exists an
alternative interpretation of quantum mechnanics in terms of
{\it consistent histories} [6] instead of von Neumann measurements.
In 1993, Paz and Zurek [7] suggested that the successive diagonalizations
(6-9) and(10-12) of the density operator led to an {\it exact} consistent set
of quantum histories for a Markovian (open or sub-) system. From that, it is
straightforward to see the continuous diagonalization, too, leads
to exact consistent histories. Actually, the orthogonal jump
process (14,15), defined uniquely for any Markovian (sub-) system,
generates exact consistent histories [3]. These histories may be {\it the}
classical content of the given quantum dynamics.

A possible lesson is that early results achieved in the frames of the
standard von Neumann measurement theory turn to be crucial if transformed
into the language of new interpretations. The old (von Neumann) and
the new (consistent history) languages both tell us the same
interpretational problems and thus wait for common solutions [8].
\bigskip

I am deeply indebted to the organizers of the Symposium for inviting me
to participate and to give this talk.
This work was supported by the Hungarian Scientific Research
Fund under Grant OTKA 1822/1991.
\bigskip\bigskip
\parindent=0in
{\bf REFERENCES}\hfill
\bigskip
1. L. Di\'osi, {\it Phys. Lett.} {\bf 112A}, 288 (1985).\hfill\break
2. L. Di\'osi, {\it Phys. Lett.} {\bf 114A}, 451 (1986).\hfill\break
3. L. Di\'osi, {\it Phys. Lett.} {\bf 185A}, 5 (1994).\hfill\break
4. G. Lindblad, {\it Commun. Math. Phys.} {\bf 48}, 119 (1976).
                                                \hfill\break
5. J. Dalibard, Y. Castin, and K. M\o lmer,
                     {\it Phys. Rev. Lett.} {\bf 68}, 580 (1992).
                                                \hfill\break
6. R. B. Griffiths, {\it J. Stat. Phys.} {\bf 36}, 219 (1984).
                                                \hfill\break
7. J. P. Paz and W. Zurek,
                     {\it Phys. Rev.} {\bf D48}, 2728 (1993).
                                                \hfill\break
8. L. Di\'osi, {\it Phys. Lett.} {\bf 280B}, 71 (1992).
                                                \hfill\break
\vfill
\end